# DIMENSIONS AND ISSUES OF MOBILE AGENT TECHNOLOGY

Yashpal Singh[1] , Kapil Gulati[2] and S Niranjan[3]

[1]P.hD, Scholar , Mewar University ,Rajasthan
yashpalsingh009@gmail.com
[2]Lecturer, BITS college of Engg,Bhiwani, Haryana
12.gulati@gmail.com
[3]Professor,PDM College of Engg , Bahadurgarh
niranjan.hig41@gmail.com

## ABSTRACT

*Mobile Agent is a type of software system which acts "intelligently" on one's behalf with the feature of autonomy, learning ability and most importantly mobility. Now mobile agents are gaining interest in the research community. In this article mobile agents will be addressed as tools for mobile computing. Mobile agents have been used in applications ranging from network management to information management. We present mobile agent concept, characteristics, classification, need, applications and technical constraints in the mobile technology. We also provide a brief case study about how mobile agent is used for information retrieval.*

## 1. INTRODUCTION

The rapid development of internet technology has led to an enormous increase in the information-access available as a result of sustained research. This makes researchers to develop systems and tools those can efficiently find, gather and retrieve this information. In this scenario, Mobile agent technology has been proposed as a model to cope up with the requirements of wide-distributed applications. A Mobile Agent is a type of software system which acts "intelligently" on one's behalf with the feature of autonomy, learning ability and most importantly mobility. More specifically, Mobile agent is a process that can transport or migrate its state from one environment to another with its data intact and is capable of performing appropriately in the new environment [1]. When a Mobile Agent moves, it saves its own state and transport this saved state to the new host and then resumes execution from the saved state. This obviously saves the network bandwidth especially in a wireless environment because once the Mobile agent is migrated; the connection between the networks is disconnected.

The remainder of this work is organized as follows: section 2 explains the Mobile Agent Technology concept, section 3 presents the characteristics of Mobile Agents. Based on characteristics, types of mobile agents have been explained in section 4. Section 5 explains the need for Mobile agent technology. The applications of Mobile agent technology are explained in section 6. Section 7 explains the case study for information retrieval from electronic calendars for multiparty event scheduling. The technical constraints in the widespread development of mobile agent technology are presented in section 8, followed by section 9 that concludes our presentation

## 2. MOBILE AGENT TECHNOLOGY CONCEPT AND TERMINOLOGY

The mobile agent in its promising paradigm provides a new means of communication amongst the network nodes. Mobile agents have been evolved from the Mobile-Code approach. Further works extended the approach to mobile –object concept in which the object (code and data) is moved from one machine to another. It will easily replace Client/Server model in future. The mobile agent has further extended this concept by moving code, data and state from one computing environment to other. Mobile agents execute at one machine, move with their state to another machine and resume execution at that new machine. These codes and objects are moved by the external command while mobile agents moved autonomously.

Before the advent of Mobile agents, the communication between the client and server is achieved by different approaches such as message passing, Remote Procedure Call (RPC) and Remote Evaluation (REV). In RPC method, the procedure resides in the server and client sends a data to the procedure that will be executed there, finally the result is back to the client. In REV approach which is different from the RPC, the procedure itself will be sent and the desired result is returned to the client. A Client/Server model is that in which server provides services to the client. The request for services is made by the client through a communication channel whether it is wired or wireless. So, when a client needs a service, it usually sends a request message to the server as shown in fig. 1. In case, the server does not have resources to satisfy the request made by the client, the client sends request to other server having the needed resource to satisfy the client that usually increase the inefficient use of network bandwidth. This also increases the network traffic and causes delays due to the involvement of more servers. These factors prohibit the widespread use of this model in a mobile device, because disconnection is frequent in the wireless environment.

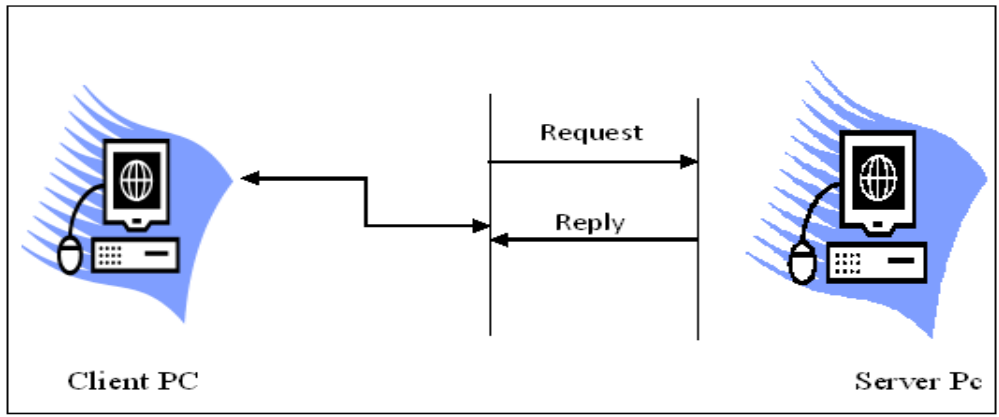

Fig.1 Client/ Server Model

Mobile agent provides solution for this mobile device because they do not depend on the server operation. Once the mobile agent has migrated, the connection between the client and server is disconnected, later when mobile agent finishes its job at the server, then it will reconnect to the client or host with the result shown in fig.2. This clearly saves the network bandwidth especially in the wireless environment where disconnection is frequent and bandwidth play a major role.

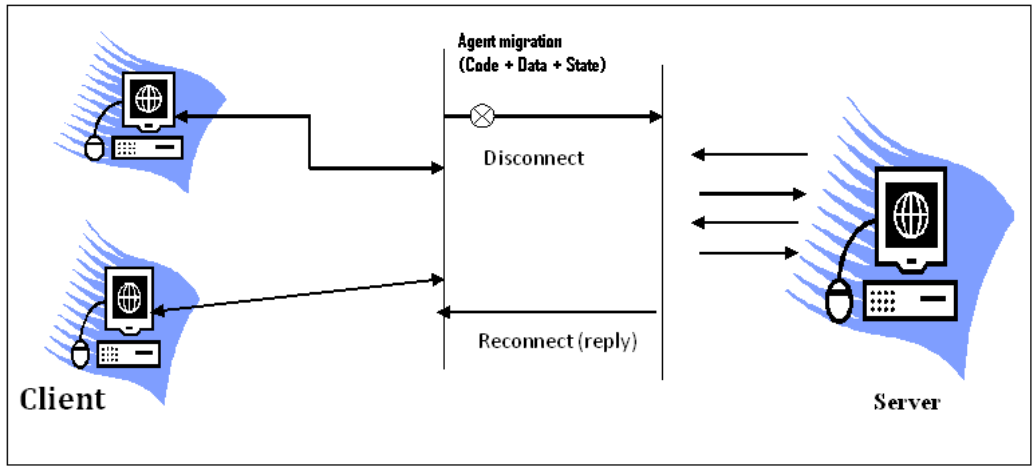

Fig 2 Mobile Agent Model

## 3. MOBILE AGENT CHARACTERISTICS

Mobile agents have different kinds of characteristics [5, 6]. They should be.

- **Autonomous**: An agent is able to take initiative and exercise a non-trivial degree of control over its own actions.
- **Interactive:** means Mobile Agents should communicate with other agents and their environment. In addition, mobility is the most important property in the Mobile Agent concept, where agent migrated from one node to another within the same environment or in different environment.
- **Coordinative;** means perform data transfer with other agents in a given environment.
- **Proxy:** Mobile agents may act on behalf of someone, so they should have certain degree of autonomy.
- **Ragged**: Mobile Agents should have the ability to deal with the errors whenever occurred.
- **Proactive**: means they should be goal oriented.
- **Cooperative:** means coordinate with other agents to achieve a common goal. Mobile Agents should have the capability of learning the current environment and modify its behaviour based on this information.
- **Intelligent**: means Mobile Agent should be too smart in order to act efficiently.

Based on these characteristics, numbers of agents have been proposed by the researchers. It is not required that agents have all these properties .This is determined by the purpose that agents have to achieve. For this we consider an example of Microsoft's software agent [7] which does not need to be mobile and cooperative but it must be smart and proactive.

## 4. CLASSIFICATION OF AGENTS

Number of agents have been built based on their purposes. Mobility is the core property in case of all.

- **COLLABORATIVE AGENT:-**An agent which collaborates with other agents to carry out an intended task. Other agents in this category can be Reactive agents, collaborative

Agents or intelligent deliberative agents. Each has a degree of expertise about some area and calls upon the expertise of other agents in the areas where it lacks expertise.

- **INTERFACE AGENT**: An *interface agent* can be considered as a program that can also affect the objects in a direct manipulation interface, but without explicit instruction from the user. The interface agent reads input that the user presents to the interface, and it can make changes to the objects the user sees on the screen, though not necessarily one-to-one with user actions. The agent may observe many user inputs, over a long period of time, before deciding to take a single action, or a single user input may launch a series of actions on the part of the agent, again, possibly over an extended period of time. The essential characteristics of an interface agent include responsiveness, competence and accessibility.

- **INFORMATION AGENT:** Information agents are special kind of so-called intelligent software agents. Software agent technology originating from distributed artificial intelligence is inherently interdisciplinary. Thus, the notion of agency is quite broadly used in literature; it might rather be seen as a tool for analyzing systems, not an absolute characterization that divides the world into agents and non-agents. However, intelligent agents are commonly assumed to exhibit autonomous behavior determined by its pro-activeness, means taking the initiative to satisfy given design objectives and exhibit goal-directed behavior, reactive or deliberative actions, means perceiving the environment and timely change management to meet given design objectives, and social in groups with other agents and/or human users when needed. It depends on the concrete application domain and views on potential solution for a particular problem what an intelligent agent in practice is supposed to do.

- **REACTIVE AGENT:** Capable of maintaining an ongoing interaction with the environment, and responding in a timely fashion to changes that occur in it. Note that the term is now widely used to mean a system that includes no symbolic representation or reasoning: such an AGENT does not reflect on the long-term effects of its actions, and does not consider the co-ordination of activity with other agents. Thus, a REACTIVE AGENT will always respond in a timely fashion to external stimulus.

- **SMART AGENT:** new forms of software agent that interface with other agents forming an artificial Intelligence system. The acronym" SMART" stands for "System for Managing Agents in Real Time". This is a bit of misnomer because the agents manage themselves and each other by agreeing to become part of the collective whole. SMART AGENTS work together within a smart system, to perform smaller pieces of larger programming tasks so that the combined collective can achieve great things with relatively simple programming building blocks. The key concept of SMART is that each individual agent does not have to be intelligent. By working together in a smart way, the agent forms a type of emergent intelligence that may appear to exhibit intelligence.

- **INTELLIGENT AGENT:** have been around for a long time in various forms. The term intelligent agent can refer to any agent that exhibits some amount of intelligence and there is no requirement that the agent have the ability to work with other agents.

- **COGNITIVE AGENT:** is a software agent that is also an intelligent agent which performs a task with minimum specific directions from the user. It evolves from the concept of virtual personal assistant, a cognitive assistant that learns and organizes.

- An Agent moves from one site to another by meeting with the **LOCAL REXEC AGENT**. The REXEC Agent expects to find two folders in the briefcase; one is HOST folder that is site, where execution is to be moved and a CONTACT folder that is agent to be executed at that site. The CONTACT folder contains the name of an agent which is a shell or a compiler. This agent would expect to find a CODE folder in the briefcase that contains the source code for an agent which it would then translate and execute .This scheme allows an agent to move to a destination site having different machine language.

- One is **COURIER AGENT** which transfers a folder to a specified agent on a specific machine.

- **DIFFUSION AGENT**, which executes an agent locally and then creates a clone of itself at every site. Scheduling allows the enforcement of policies that tells when and where an agent is executed .

- This scheduling is implemented by **BROKER AGENT**, which maintains the database of service providers .An Agent needs a given service always consult a broker to identify which agent provides that service.

- **REAR GUARD AGENT,** it is possible that sites in a computer network will fail. When such a failure occurs, agents at that site will no longer be continuing execution. To solve this problem a REAR GUARD AGENT can be deployed so that the execution can proceed. This agent follows the execution process which moves from one site to another. When an agent fails, this REAR GUARD AGENT launches a new agent. On resumption, this REAR GUARD Agent terminates its own action.

## 5. THE NEED FOR MOBILE AGENT TECHNOLOGY

The Mobile Agent has its applications in many areas including network management, mobile computing, information management, web services, remote software management and others. Mobile Agents enhance the performance in these areas by providing the following services: Efficiency and reduction of network traffic: Mobile agents consume fewer network resources since they move the computation to the data rather than the data to the computation. Also mobile agents can package up a conversation and ship it to a destination host, where the interactions can take place locally, so network traffic is reduced (figure 3).

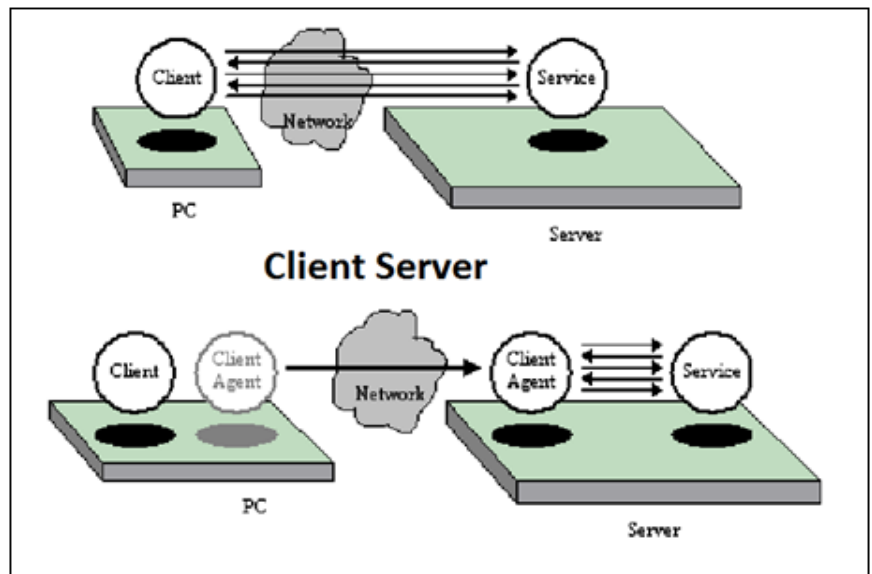

Fig 3

Asynchronous autonomous interaction: Tasks can be encoded into mobile agents and then dispatched. The mobile agent can operate asynchronously and independent of the sending program. Interaction with real-time entities: Real-time entities require immediate responses to changes in their environment. Controlling these entities from across a potentially large network will incur significant latencies. Mobile agents offer an alternative to save network latency. Local processing of data: Dealing with vast volumes of data when the data is stored at remote locations, the processing of data over the network is inefficient. Mobile agents allow the processing to be performed locally, instead of transmitting the data over a network. Support for heterogeneous environments: Both the computers and networks on which a mobile agent system is built are heterogeneous in character. As mobile agent systems are generally computer and network independent, they support transparent operation. Convenient development paradigm: The design and construction of distributed systems can be made easier by the use of mobile agents. Mobile agents are inherently distributed in nature and hence are natural candidates for such systems.

There are three basic needs for Mobile Agents to achieve these goals, the Mobile Agent Program, Mobile Agent Platforms and Mobile Agent Creator. Today the mobility is performed by different coding methods. So, the conventional Programming Languages cannot be applied in implementation. However, implementation through java due its independent execution environment is somehow being managed to build distributed applications. The following operation: Creation, Cloning, Dispatching or Migration, Retraction, Activation, Deactivation and finally Disposal have been carried out through java [8, 9]. These operations constitute the mobile agent life cycle. Creation is the first phase in the mobile agent life cycle. Whenever a request is made to the mobile agent, a mobile agent instance is created which means a desired parameter is equipped with the mobile agent to achieve its goal before any further work is done. Cloning refers to creating a copy of the original mobile agent object. This operation is used when the need for another agent with the same feature arises. The migration or Dispatching is used for moving the agent from one node to another by specifying address of the destination. This migration is of two types. One is strong Migration in which the mobile agent itself, its data and its state move. Second one is weak Migration which includes mobile agent itself and its data. The Retraction function is used whenever agent's source node required that its agent returned to the original host after completion of its job. To start and stop of mobile agent is done by the Activation and Deactivation operation. Finally Disposal operation is done at the end of the mobile agent life cycle. Fig.4 shows the mobile agent life cycle (operation).

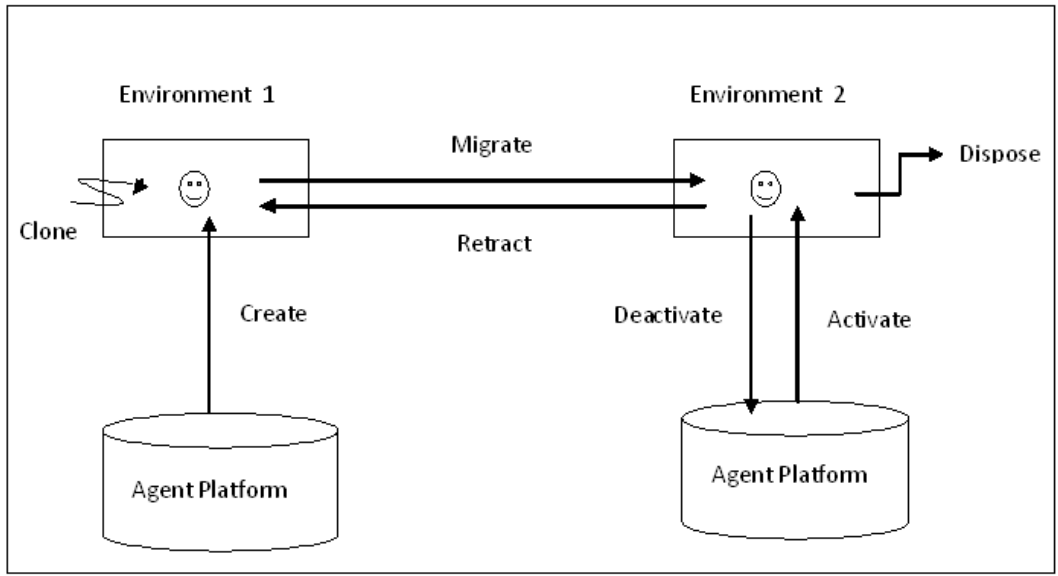

Fig. 4 Mobile Agent life cycle

The second requirement is the mobile agent platform or execution environment. Mobile agent platform must be implemented and exists to run the mobile agent application. Mobile agent platform must have some special characteristics so that the host may know how to deal with the incoming agents and provides the environment to those agents so that they can achieve their goals. For this, the requirement should be platform independence, authentication, secure execution, dynamic class loading, network connectivity and resources control. These requirements should be provided by the mobile agent platform. Over the past few years, numbers of industrial mobile agent platform have been proposed which provides these resources such as Telescript[14], Aglets[8], Concordia[4] and Voyager[10].

The third requirement is the mobile agent creator means developers who developed the mobile agent system based on two main models. The first model which is just as an extension to the operating system function, controls the mobile agent life cycle and provides one platform per host. This model does not give much flexibility because each agent has operated based on their allocated platform and does not operate independently. The second model is compound based model which separates the platform from the host. This separation helps developers to develop a wide range of mobile agent applications because now agents operate autonomously [11].

## 6. APPLICATIONS OF MOBILE AGENT TECHNOLOGY

Mobile agent technology has been used in many areas from network management task to information management. Mobile agents have significantly used in the wireless environment because it supports the disconnected mode. As the mobility has been migrated from the PDA (Personal Digital Assistant) to the network, the PDA could be disconnected and when the mobile agent finishes its job then they can reconnect in the network with the desired result from the agents. This gives advantage over conventional communication methods such as client/server model, RPC etc.

Mobile agent technology is frequently used in other applications such as data warehouse, software updates, information management tasks such as searching for information, information filtering, information monitoring etc. Mobile agent technology has been also applied in M-commerce for information retrieval regarding the lower price of any particular item. In this, mobile agent have been issued by the PDA and disconnected from the network. This agent will then roams from one node to another on the internet and compare the price. When the mobile agent finds the cheapest result, it will then return to the PDA. This concept of information retrieval is further explained by considering a case study of multiparty event scheduling.

## 7. CASE STUDY

Our case study is on information retrieval from electronic calendars for multiparty event scheduling [13]. Many events require the participation of several parties. Before knowing the date when most (not all) participants will be available is all depends on their schedules. But to identify this date is always a typical task when number of targeted participants is large. Today, electronic agendas are stored on server. An application can access them and retrieve information. In our case study, we dispatch a mobile agent in the network instead of using client/server paradigm. In this approach, the agent visits the server, access the agendas, retrieve the information and identifies the date. This is explain in figure 5.

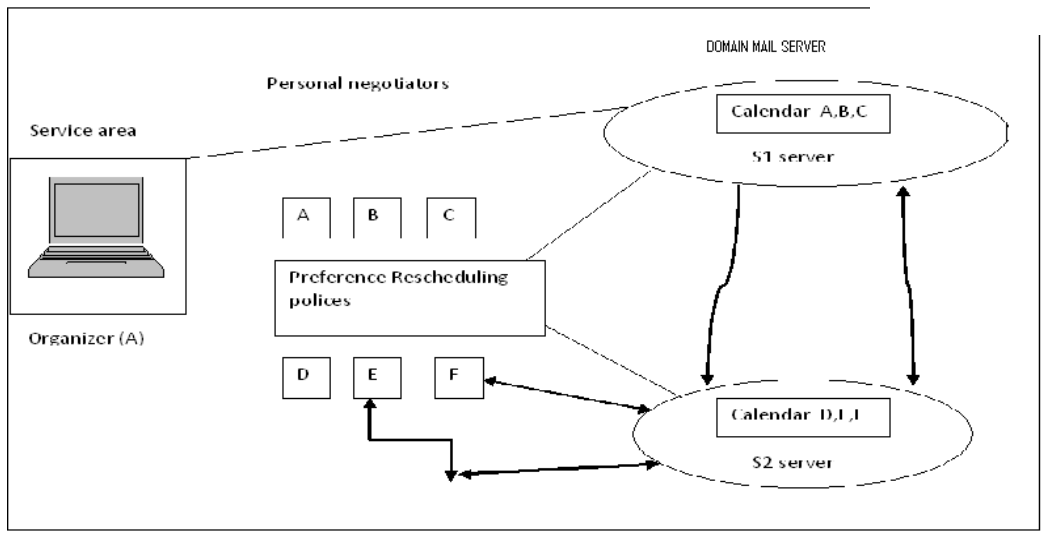

Fig 5. Mobile Agent Scheduler Scenario

The conference has been organized by A and the targeted participants are A itself, B, C, D, E and F. In this A, B and C are resides in S1 domain and D, E and F resides in S2 domain. There is one server per domain. Each participant has their personal agent who can work on behalf of them. The condition for conference is that, it will be held only if at least four participant are available and it will take place in a given month (specified).

The organizer A gives the name and e-mail address of targeted participant along with the month when the event takes place to his own agent. The agent connect itself to S1 and retrieves the calendars of A, B and C for a given month and process the information to find the date when all three participant will be available. It could be few days. After that the agent moves to S2 and retrieve the information of D, E and F and compare with the dates of A, B and C. The agent finds that two dates are suitable for F but fails to find the date with D and E. Then, this scheduler agent proposes these dates to the agents of D and E and finds that E can reschedule his event but D could not. The agent finally sends a notification to the A, B, C, E and F because four participant are available, which is necessary conditions for events and finally event will take place. This surely increases the performance of the system.

## 8. TECHNICAL CONSTRAINTS

The mobile agent is relatively a new technology and provides new way of communication amongst network nodes. There are number of successful mobile agent applications but still there are some barriers preventing the wide spread of this technology. There are so many reasons; one of them is lack of standard in both software and hardware products e.g programming language, protocols and devices. Researchers are working on these standards so that developers can build a powerful applications based on mobile agent technology. The lack of understanding about what the mobile agent should perform is another contributing problem. The infrastructure is not suitable for implementing the mobile agent's technology [12].

There are several issues such as security, privacy, trust, integrity etc. This can explain by considering a case where mobile agents are used in E- commerce or M- commerce for transaction over the internet. It means, Mobile Agent will access the user profile which contains some sensitive information about the users, which may be shared with other agents in the working environment .This information may be modified during transactions . So, trust and privacy lost. That's why we need some methodology so that information could be send without changes its contents or we can say that, Mobile Agents need to be protected against hosts and hosts need to be protected against mobile agents [2,3].

Other issues such as today's market are based on conventional client/server model. So, we have to rethink in the design so that mobile Agent technology could be employed with ease. To summarise, the table below briefly explains the implication of using Mobile agents:

The Technical Implications of Mobile Agents

| Technical Issue | Implication for Mobile Agents |
|---|---|
| Bandwidth | MAs conserve bandwidth, especially for networks which have low bandwidth capacity(e.g wireless network). By replacing continous communication with an agent directly at the point of information generation, the bandwidth use can be reduced. Instead of sending dozen or even hundreds of queries across the network, sending one agent on a single request the agent can manage this process locally at the remote side. |
| Fault-tolerance | MAs can act or respond on errors that may be encountered within their contexts because of their adaptive and ragged attributes. |
| Flexibility | MAs can give greater flexibility, because new tasks and codes can be added to the system without the need for a fixed code-base. |
| Interaction | MAs enable new type of interaction, such as negotiating agents that travel to server site seeking for the best deal such as comparing prices(e.g e-commerce application). |
| Protocols | MAs are able to move to remote hosts in order to establish channels based on proprietary protocols. |
| Scalability | MAs can carry out their function well (without disruption) when the host system or environment changes in size or volume in order to meet a new user's need. |
| Self-contained tasks | MAs can carry out tasks which require variable degrees of independence such as network management, software updates etc. |
| Weak coverage | MAs fit perfectly into a disconnected environment where the signal coverage is frequently lost (being disconnected): MAs will then migrate from one node to another when the coverage becomes available. |

## 9. CONCLUSION

Mobile agent technology provides a new way of communication over heterogonous network environment. A number of advantages have been proposed and identified which includes: efficiency and reduction of network traffic, asynchronous autonomous interaction, interaction with real-time entities, local processing of data, support for heterogeneous environment and having robust and fault-tolerant behaviour. However, the security, infrastructure and standardising issues still represent significant constraints. The main thing which we concluded from our findings and investigation is that mobile agent technology has the potential in increasing the performance of networks as well as for software adopting mobile agents. Due to its nature of being a futuristic technology from the programming environment perspective a lot of work is still required before the average programmer can build applications based on the mobile agent technology paradigm with ease.

## Authors

**Mr. Yashpal Ssingh** obtained his MCA(2006) from G.J.U. University, M.Tech(CSE-2009) C.D.L.U University and Ph.D (CSE) Pursuing From Mewar University ,Rajasthan ,India. He has attended various national, International seminars, conferences and presented research papers on Artificial Intelligence ,Mobile Agent and Multi-Agent Technology.

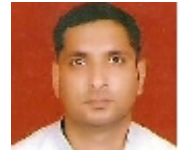

**Niranjan S** was born on 4[th] April 1955 in India. Graduated from University College of Engineering, Sambalpur University in 1978 in Electronics & Telecommunication Engineering, worked as asst Engineer under State Govt and Ex. Engineer under State Electricity Board working for Power line Carrier Communication Systems, Protective relaying, telemetry and LFC Systems. Master's Degree from IIT Kharagpur in 1987 in Computer Engineering. Worked in the areas of Parallel Processing.Performance characterization of parallel Programs under variable and uniprocessor environments, worked towards the development of a machine and hand printed Oriya Character Recognition System. Ph.D . from Utkal University. Area of work is Study and issues of mobile Computing Algorithms. He has also worked in the areas of Load Frequency Control in deregulated Scenarios and study of various non linear models of interconnected power Systems including GRC and other stochastic conditions. His current research areas include ECC based cryptographic applications, mobile ad-hoc sensor networks.

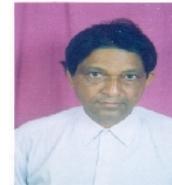

**Mr. Kamal Deep** obtained his BTech(2007) from KUK. University, M.Tech(-2011) M.D.U University and working as a Assistant Professor in BRCM bhel ,haryana ,India.

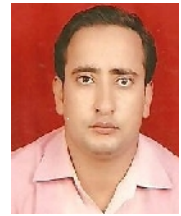